# Molecular Dynamics Study of the Primary Ferrofluid Aggregate Formation

**B. M. Tanygin[a], V. F. Kovalenko [a], M. V. Petrychuk [a], S. A. Dzyan [a]**

[a] Radiophysics Department, Taras Shevchenko Kyiv National University, 4G, Acad. Glushkov Ave., Kyiv, Ukraine, UA-03127

*Corresponding author:* B.M. Tanygin, Radiophysics Department, Taras Shevchenko Kyiv National University, 4G, Acad. Glushkov Ave., Kyiv, Ukraine, UA-03127.

*E-mail*: b.m.tanygin@gmail.com

*Phone*: +380-68-394-05-52

**Abstract.**

Investigations of the phase transitions and self-organization in the magnetic aggregates are of the fundamental and applied interest. The long-range ordering structures described in the Tománek's systematization (*M. Yoon, and D. Tománek, 2010 [1]*) are not yet obtained in the direct molecular dynamics simulations. The resulted structures usually are the linear chains or circles, or, else, amorphous (liquid) formations. In the present work, it was shown, that the thermodynamically equilibrium primary ferrofluid aggregate has either the long-range ordered or liquid phase. Due to the unknown steric layer force and other model idealizations, the clear experimental verification of the real equilibrium phase is still required. The predicted long-range ordered (crystallized) phase produces the faceting shape of the primary ferrofluid aggregate, which can be recognized experimentally. The medical (antiviral) application of the crystallized aggregates has been suggested. Dynamic formation of all observed ferrofluid nanostructures conforms to the Tománek's systematization.

**Keywords:** ferrofluid, primary aggregate, nanostructure, long-range ordered phase, faceting shape, molecular dynamics simulation.

## 1. Introduction

The colloidal suspension with nanoparticles carrying electric or magnetic dipole moments [1,2] are systems with a wide range of applications [3]. The important application is a cancer treatment: magnetically targeted drug delivery and magnetothermal therapy [4]. The development of the ferrofluid based photon crystal has been suggested in the Ref. [5]. The suspended and aggregated phases are two major states of the ferrofluid matter. The former phase has the superparamagnetic properties, leading to the effective magnetic relaxation for the magnetothermal therapy: low coercitivity and high magnetic susceptibility requirements. The aggregated phase is more complex and less studied due to variety of possible magnetic moment and spatial position orderings of the nanoparticles inside the ferrofluid aggregate. This phase is neither superparamagnetic nor ferromagnetic. Consequently, the investigations of the phase transitions and self-organization in the magnetic aggregates are of the fundamental interest.

The equilibrium structures of the ferrofluid aggregates are produced by the competition between different interaction types. The anisotropic dipole–dipole interaction generates the linear chains, circle structures, and their complex combinations [6-8]. Oppositely, the isotropic attractive forces [6,7] favor the structures with the compact packing. The Brownian motion favors the superparamagnetic “gas phase” of the magnetic particles. If the intensity of the Brownian motion of the particles is not enough to produce the liquid phase (microdrop aggregates) or amorphous solid phase then the long-range ordered compact packing corresponds to the minimum of the free energy. Hereinafter, we use terms “phase”, “solid”, “liquid”, “crystal”, etc. to describe the type of the nanoparticles spatial position ordering inside the ferrofluid aggregate. We use term “crystallized aggregate” to determine the long-range ordered phase of the aggregate. It is well-known, that the difference between liquid and amorphous phases has quantitative nature. This conventional quantitative threshold between these phase types is not established in scope of the present investigations. Hereinafter, we will use only term “liquid phase”.

Due to the good experimental accessibility, the colloidal suspensions are interesting model systems for studies of crystal nucleation and growth [9]. The different phase transitions are possible in the ferrofluid aggregates including the gas-liquid phase transition behavior [10]. It was experimentally shown [11,12], that after creation of the ferromagnetic $Fe_3O_4$ nanoparticles inside the carrier liquid, they combines into the isolated assemblies (primary aggregates [11,12]). Primary aggregate has the confined magnetic flux and small total magnetic moment with order of magnitude of the single nanoparticle magnetic moment [11,12]. The variety of possible shapes, structures and sizes of solid [13-19] primary aggregates is expected. The internal structure and shape details of such primary aggregates were not directly observed.

Generally, the phase evolution can be described as the following transitions: chemical synthesis of the nanoparticles → superparamagnetic gas phase → linear chains and/or circles → connected ring assemblies, coils, tubes and scrolls [1] → liquid phase aggregates [10,18,19,22,23] → crystallized primary aggregate (chain assembly [1])→ secondary aggregates (filiform [13-17], rod-shaped [11,12], and dumbell-like [12] aggregates). Each of these phases can be equilibrium against the special experimental conditions. For instance, the regular ferrofluid in case of the magnetothermal therapy application has the superparamagnetic gas phase.

The primary aggregates have been investigated on the basis of the continuous model [1], where long-range ordering of the nanoparticles was defined in the model. The compact packing of the equal size spheres corresponds to the lattices with the highest average density: face-centered cubic and hexagonal close-packed lattice [20]. These phases can coexist in the crystallized aggregates. The phenomenological theory of the aggregates [1] can be verified using the microscopic molecular dynamics study. Alternatively, it can be made on the basis of the thermodynamic theory. Most molecular dynamics simulations usually focus on the systems with periodical boundary conditions [6,7]. The long-range ordering predicted by the theory [1] is not obtained in these simulations. The resulted structures usually are the linear chains or circles [6-8,21], or, else, liquid formations [22,23] obtained after coagulation of the linear chains.

The assumption, which initiates the present investigations, is that these liquid formations can represent metastable or intermediate phase, which finally transforms to the crystallized phase after long enough simulation time. The purpose of this work is building the simulation method and investigation of the long-range ordering phase transition in the ferrofluid primary aggregate.

## 2. Calculation method

### 2.1 Model forces

The simulation method includes the modeling core and real time 3D graphical support. The technical implementation is publically accessible [24]. The calculations have been implemented in the SI units.

As in the phenomenological theory [1], the present molecular dynamics simulation is dedicated to the finite number of particles without periodic conditions. The spherical ferrofluid nanoparticles are single-domain. The particles size and magnetic moment dispersion are neglected. The steric layer (stabilizing surfactant) thickness is $\Delta R$. The total diameter is $\sigma = 2(R + \Delta R)$, where $R$ is a nanoparticle radius.

Starting conditions of the simulation is a random distribution of the nanoparticles magnetic moments $\mathbf{m}_i$ and positions $\mathbf{r}_i$ in the 3D space $560 \times 560 \times 560$ nm, where minimal starting distance between nanoparticles is $\min \mathbf{r}_{ij} = \sigma$. In order to minimize the time of the relaxation and to maximize the accuracy, the hard walls boundary conditions have been selected. Such model is proper for isolated primary aggregates (which have the confined magnetic flux) formation in contrast to the formation of the secondary aggregates [11,12].

The total force acting on the $i$-th nanoparticle is given by:

$$\mathbf{F}_i = \mathbf{F}_i^f + \mathbf{F}_i^g + \sum_j \left( \mathbf{F}_{ij}^{dip} + \mathbf{F}_{ij}^s \right) \quad (1)$$

where the random force related to the Brownian motion was not yet included. It will be taken into account later on the statistical basis. Here, the $\mathbf{F}_i^g$ is a sum of the gravity and the buoyancy force. The $\mathbf{F}_i^f = -\lambda_f \mathbf{v}_i$ is a viscous friction force, where $\mathbf{v}_i$ is the $i$-th nanoparticle speed vector, $\lambda_f = 3\pi\eta\sigma$, and $\eta$ is the water viscosity at the $T = 293$ K and normal pressure. The force $\mathbf{F}_{ij}^{dip}$ is derived from the dipole-dipole potential:

$$\mathbf{F}_{ij}^{dip} = \frac{3\mu_0}{4\pi \mathbf{r}_{ij}^5}\left[(\mathbf{m}_j\mathbf{r}_{ij})\mathbf{m}_i + (\mathbf{m}_i\mathbf{r}_{ij})\mathbf{m}_j + (\mathbf{m}_i\mathbf{m}_j)\mathbf{r}_{ij} - \frac{5(\mathbf{m}_i\mathbf{r}_{ij})(\mathbf{m}_j\mathbf{r}_{ij})}{\mathbf{r}_{ij}^2}\mathbf{r}_{ij}\right], \tag{2}$$

where $\mathbf{r}_{ij} = \mathbf{r}_i - \mathbf{r}_j$.

The $\mathbf{F}_{ij}^{s}$ is the $i$-th and $j$-th nanoparticle steric layers overlapping force. It is still not obtained, which kind of $\mathbf{F}_{ij}^{s}$ is most suitable to model real ferrofluids [6]. The logarithmic [23], exponential [21], and Lennard-Jones [6] dependence have been utilized. Common feature of these models is the isotropic and short-range nature of the force. Exponential and Lennard-Jones potentials have high gradients. Therefore, they require special technique of simulations or usage of the high time quantization. Oppositely, we will design the isotropic and short-range model force without high-power or exponential dependences, which drastically simplify simulation process due to the weaker time quantization level requirements.

The repulsion force can have form which guarantees avoiding the violation of the nanoparticle hardness. Consequently, the repulsion force existence is determined by the cutoff $\theta[\sigma - |\mathbf{r}_{ij}|]$, where $\theta[x]$ is a Heaviside step function. The special form of this potential should not affect the equilibrium state. In order to balance the dipole-dipole attractive forces, the repulsion force dependence of the type $\mathbf{r}_{ij}/\mathbf{r}_{ij}^5$ is a minimal power, which can be selected. The same dependence can be selected for the steric layers attractive force. In contrast to the dipole-dipole interaction, the $\mathbf{F}_{ij}^{s}$ cannot generate the long-range effects. Thus, the model cutoff $\theta[3\sigma/2 - |\mathbf{r}_{ij}|]$) of steric layers attractive force starting from some distance (we selected $3\sigma/2$) cannot affect the simulation model adequacy. Thus, in our model the steric layers attractive force is the force of the sticking nature.

Consequently, the short range approximation of the steric layers interaction potentials can be selected in the following form:

$$\mathbf{F}_{ij}^{s} = \lambda_r \frac{\mathbf{r}_{ij}}{\mathbf{r}_{ij}^5}\theta[\sigma - |\mathbf{r}_{ij}|] - \lambda_a \frac{\mathbf{r}_{ij}}{\mathbf{r}_{ij}^5}\theta[|\mathbf{r}_{ij}| - \sigma]\theta\left[\frac{3\sigma}{2} - |\mathbf{r}_{ij}|\right], \tag{3}$$

where repulsion and attraction constants were selected as: $\lambda_r = 15\mu_0 m_0^2/(4\pi)$ and $\lambda_a = 3C_a\mu_0 m_0^2/(4\pi)$, $m_0 = |\mathbf{m}_i|$. Here, the $\lambda_r$ form was selected to guarantee the nanoparticle hardness against the strong dipole-dipole attraction (2). The $\lambda_a$ is a function of the phenomenological constant $C_a$, which determines the value of the steric layers attractive force against dipole-dipole attractive force. Selecting of this non-zero parameter does not affect equilibrium ferrofluid structure. It affects only the simulation time required for the obtaining of the final equilibrium structure.

The rotational motion of the nanoparticle was not explicitly taken into account in the present simulation. Real alignment of the magnetic moment of the magnetic ferromagnetic nanoparticle along the self-confidence field of the system is provided via the relaxation oscillations. Order of magnitude of such damping time is given by nanoparticle moment of inertia $I$ and rotational friction constant $\Gamma_R$ relation. The value of the $\Gamma_R$ does not affect equilibrium state. In the Ref. [6], this value has been selected on the basis of the simulation optimization. The model assumption of the present simulation is the following relation: $I/\Gamma_R < \Delta t$, where $\Delta t$ is the time quantization step in the difference schema. It means that the rotation and rotational damping of particles is considered as fast enough. In this case, nanoparticle aligns its magnetic moment along magnetic field $\mathbf{B}(\mathbf{r}_i)$. On *each* step of the dynamic problem solving, we should solve the micromagnetic problem using the LaBonte difference schema [25]. First, each magnetic moment $\mathbf{m}_i$ is aligned along the field $\mathbf{B}(\mathbf{r}_i)$ produced by all magnetic dipoles. Then, new field is calculated and schema is repeated until accuracy is reached. The accuracy threshold is determined via the new field and previous magnetic moment reflection angle:

$$\max \arccos\left[\frac{\left(\mathbf{m}_i \mathbf{B}(\mathbf{r}_i)\right)}{m_0|\mathbf{B}(\mathbf{r}_i)|}\right] < \epsilon, \qquad (4)$$

where maximal angle is selected among all nanoparticles. We chose the following threshold: $\epsilon = 0.2$ rad.

The coherent or incoherent magnetization switching are the additional and/or alternative mechanisms of the magnetization alignment along the $\mathbf{B}(\mathbf{r}_i)$. The model assumption regarding such alignment time and $\Delta t$ value is the same as described above.

## 2.2 Finite-difference scheme of the molecular dynamics

It is naturally, that force $\mathbf{F}_i^f(\mathbf{v}_i)$ makes the differential equation non-linear. Therefore, simulation becomes difficult. This force changes in range from zero ($\mathbf{v}_i = 0$) to finite value passing one time step. The difference schema becomes divergent. To avoid it, it is possible to solve the motion equations against unknown function $\mathbf{v}_i$ in the time interval $(t, t + \Delta t)$:

$$\mathbf{v}_i(t + \Delta t) = \frac{\tilde{\mathbf{F}}_i(t)}{\lambda_f} + \left[\mathbf{v}_i(t) - \frac{\tilde{\mathbf{F}}_i(t)}{\lambda_f}\right] e^{-\frac{\lambda_f \Delta t}{M}}, \tag{5}$$

where $M$ is a nanoparticle mass. Here, the *relative* changes of all other forces $\tilde{\mathbf{F}}_i(t) = \mathbf{F}_i(t) - \mathbf{F}_i^f(t)$ are neglectfully small at the relation: $d_m = \max|\mathbf{r}_i(t + \Delta t) - \mathbf{r}_i(t)| \ll \sigma$, where maximum is obtained among all nanoparticles. We selected threshold $d_m/\sigma < 0.2$. If this condition is violated at least for the one particle then simulation iteration over the time step is reversed and time step value is decreased: $\Delta t \to \Delta t/2$ . After 10 time steps the quantity $\Delta t$ is restored. The start value is $\Delta t = 10^{-5}$ s. The regular value selected by algorithm is $10^{-9}..10^{-7}$ s. The smallest value $10^{-9}$ s corresponds to the relation $\frac{\lambda_f \Delta t}{M} \sim 10$. It means that exponent of (5) cannot be expressed into the series $1 - \frac{\lambda_f \Delta t}{M} + o\left(\frac{\lambda_f \Delta t}{M}\right)$. Consequently, the $\mathbf{F}_i^f(\mathbf{v}_i)$ integration approach makes the simulation process at least 10..100 times faster than method of the simple finite difference calculation of the nanoparticle acceleration produced by the total force.

The integration of (5) and inclusion of the translational Brownian motion gives the following expression:

$$\mathbf{r}_i(t + \Delta t) = \mathbf{r}_i(t) + \frac{\tilde{\mathbf{F}}_i(t)\Delta t}{\lambda_f} + \frac{M}{\lambda_f}\left(\mathbf{v}_i(t) - \frac{\tilde{\mathbf{F}}_i(t)}{\lambda_f}\right)\left(1 - e^{-\frac{\lambda_f \Delta t}{M}}\right) + \delta\mathbf{r}_i(t) \tag{6}$$

Here, the Brownian motion in the potential field is determined by the pseudo random numbers generation in scope of the Boltzmann statistical distribution. The Boltzmann probability of the shift $\delta\mathbf{r}_i(t) = \sqrt{6D\Delta t}$ is given by:

$$dp = \frac{W}{2\pi(e^{W} - e^{-W})} e^{W\cos\theta} \sin\theta \, d\theta d\varphi, \quad (7)$$

where the diffusion coefficient is $D = RT/\,(N_A \lambda_f)$. Here, the normalized potential energy decrease is given by $W = |\tilde{\mathbf{F}}_i(t)|\sqrt{6D\Delta t}/k_B T$. The angle $\theta$ between $\tilde{\mathbf{F}}_i(t)$ and $\delta\mathbf{r}_i(t)$ affect the probability, i.e. random motion along the total force vector is the most probable. The angle $\varphi$ is a free rotation of $\delta\mathbf{r}_i(t)$ around the $\tilde{\mathbf{F}}_i(t)$ with uniform probability distribution. No rotational Brownian motion was included into the present model. The thermal fluctuations of the magnetization direction also were not considered.

## 3. Results and discussion

### 3.1 Phase transitions

The number of particles was selected as $N = 400$. The magnetite saturation magnetization 80 emu/g and mass density 5.05 g/cm$^3$ were selected. The radius is $R = 10$ nm. The carrier liquid is water under the temperature $T = 293$ K. The selected parameters are $\Delta R = 10$ nm (thick layer) and $C_a = 1.5$. These parameters describe the stabilizing surfactant layers interaction. However, it was determined that selection of different values of $\Delta R > 0$ and $C_a > 1$ does not change the type of the equilibrium ferrofluid structure. These parameters affect the simulation time and quantitative characteristics (e.g. size) of the final structure.

The evolution of starting superparamagnetic gas phase dynamics toward the final phase was simulated. The aggregate formation was described using the total moment of inertia and total magnetic moment dependencies (Fig. 1) for the whole time of the simulation. Minimization of the free energy of the dipole-dipole interaction leads to minimization of the total magnetic moment. In case of the superparamagnetic state with random distribution of the nanoparticles positions (initial state of present simulation) and absence of the external field, the mathematical expectation of the total magnetic moment is zero (start value at Fig. 1b). Once nanoparticles form some non-continuous structures, there is no rule which requires zero value of the total magnetic moment. Once separate chains of the nanoparticles form single continuous sample (liquid or crystallized), the micromagnetics laws (minimization of the demagnetization fields energy) of continuous magnetically ordered media lead to the minimization of the total magnetic moment again (Fig. 1b).

Different regions of the setting dependencies (Fig. 1) conform to the different structures of the ferrofluid shown at the specific time points (Fig. 2-3). The first structural transition is nanoparticles connection into the linear chains (Fig. 2a) at the time point $t = 0.2$ ms. Next stage is their locking into the circles. The circles and linear chains connect and form the ring assemblies, coils, tubes and/or scrolls [1]. The present simulation results contain combination of these structures. The ring assemblies and tubes predominates (Fig. 2b-g). Small aggregate in the right bottom corner has ring assemble structure with the total magnetic moment $3m_0$ (Fig. 2b-c). If large aggregate has no total magnetic moment then the small aggregate induces it. Finally, due to the attraction by the magnetic field gradient, the small rings assemble and large tube aggregates are merged (Fig. 2d-f) during the time range $t = 1.7 \dots 2.4$ ms.

The tube is shrinking (Fig. 2e,g) until the cavity disappears (Fig. 3a-b) at the time point $t = 4.6$ ms. The aggregate has the confined magnetic flux (Fig. 3d). The surface shape is close to the sphere (Fig. 3c). On the basis of the analysis of the bulk structure, it was derived, that there is no long-range order. This liquid state continuously transforms into the crystallized phase ending at the time point $t \cong 0.42$ s. This process of crystallization is temporally continuous as well as dynamics of any first-order phase transition. The magnetic flux of the crystallized aggregate is confined (Fig. 3f). However, shape is not spherical. Predictably, both dipole-dipole and isotropic forces favor the compact packing of spheres. The spheres occupy the 74% of the final aggregate space (limit density of compact packing [20]). Such packing has Bravais lattice, and, therefore, surface formation is a direct analogy of the crystal growth (Fig. 3e). The long-range order can be illustrated by the clearances, which are observed along the certain “crystallographic” directions (Fig. 3g). Total magnetic moment of the final aggregate has order of magnitude 1..5 $m_0$ at the total number of nanoparticles $N = 400$ (Fig. 1b).

### 3.2 Applications

The Brownian motion is able to prevent the long range ordering. Thus, real primary aggregate structure can have either liquid or crystallized phase. Defects with different dimensions (e.g. phase boundaries) are natural result of the spontaneous formation of the aggregate. The crystallized phase can be experimentally detected using the direct microscopic observation (e.g. the transmission electron microscopy [26]) of the primary aggregate shape. The experimental setup details required for the formation of the primary aggregates are described in the Refs. [11,12]. If liquid phase is glass-like one then crystallization process can be long-continued. The aggregates precipitation was observed in the experiments [11,12], which restrict number of particles. At the precipitation stage, primary aggregates with diameter $\sim 650$ nm [11,12] (number of particles ~15 000 at $R = 10$ nm and $\Delta R \ll R$) emerge and combine into the secondary (rod-shaped [11,12], and dumbell-like [12]) aggregates. Interaction between primary aggregates is an aim of further research.

Generally, the crystallized phase is theoretically possible. However, due to the unknown steric layer force (3) and others model idealizations specified above, the clear experimental verification of the real equilibrium phase is required. The viscosity (type of the carrier liquid), type of surfactant (different values $\Delta R$ and forces), nanoparticle sizes and temperature can be varied in the experiments. Increasing of the temperature, obviously, transforms crystallized aggregate into the liquid one. In a real ferrofluid the particles size and magnetic moment dispersion perturb faceting shape making it less symmetric. The observation of the crystal faceting shape of the aggregate (polygon cross-section) is a criterion of the long-range ordered phase (e.g. by the transmission electron microscopy). Also, crystallized phase (Bravais lattice nanoparticles order inside aggregates) can be detected by the UV and long wave-length X-ray diffraction. The observation of the spherical shape is a criterion of the liquid phase.

The total magnetic moment equals 0.25%-1.25% of the saturation magnetic moment. Consequently, the magnetic susceptibility of the ferrofluid aggregated phase is smaller than in case of the superparamagnetic suspended phase. Thus, higher magnetic field magnitude is required for the magnetothermal cancer therapy applications in medicine. The macroscopic mechanical properties of the ferrofluid are determined by the hardness of the primary aggregates (liquid vs. crystallized) in the suspension.

The liquid-to-crystallized phase transition can be utilized in the medical applications. The time of this crystallization process $\Delta t_c \sim 0.42$ s has been obtained in present simulation. The liquid primary aggregate of size $\sim 650$ nm is able to phagocytize biological objects in vivo (e.g. HIV virion of size $\sim 120$ nm [27] in the blood). Once it became crystallized, the aggregate become hard and can effectively conserve virion. The virion can be killed by high temperature [28] produced by the magnetic relaxation of the primary aggregate in the alternating field (same, as in case of the magnetothermal therapy [4]). Optionally, the crystallized aggregate can be removed from blood by the magnetic field gradient.

## 4. Conclusions

Evolution of all observed ferrofluid nanostructures conforms to the Tománek's systematization. The linear chains and/or circles dynamics produces formation of the ring assemblies and coils. Connection of these structures gives the tubes and scrolls. Finally liquid and crystallized aggregates have been formed. The final aggregate has substantially confined magnetic flux.

Thus, the primary ferrofluid aggregate has liquid or long-range ordered (crystallized) phase. The latter phase requires experimental verification. Sphere close packing corresponds to the Bravais lattice. The observation of the crystal faceting shape of the aggregate (polygon cross-section) is a criterion of the long-range ordered phase (e.g. by the transmission electron microscopy). The macroscopic mechanical properties of the ferrofluid are determined by the hardness of the primary aggregates (liquid vs. crystallized) in the suspension. It was shown, that liquid-to-crystallized phase transition complete in 0.42 seconds. This phase transition can be utilized in the medical applications (e.g. antiviral therapy).

## Acknowledgements

We thank D. M. Tanygina for support with a graphical design. We thank Mr. E. Slusar' and Mr. A. Kokhanovskyi for the support with OpenGL 3D visualization.

**a)**

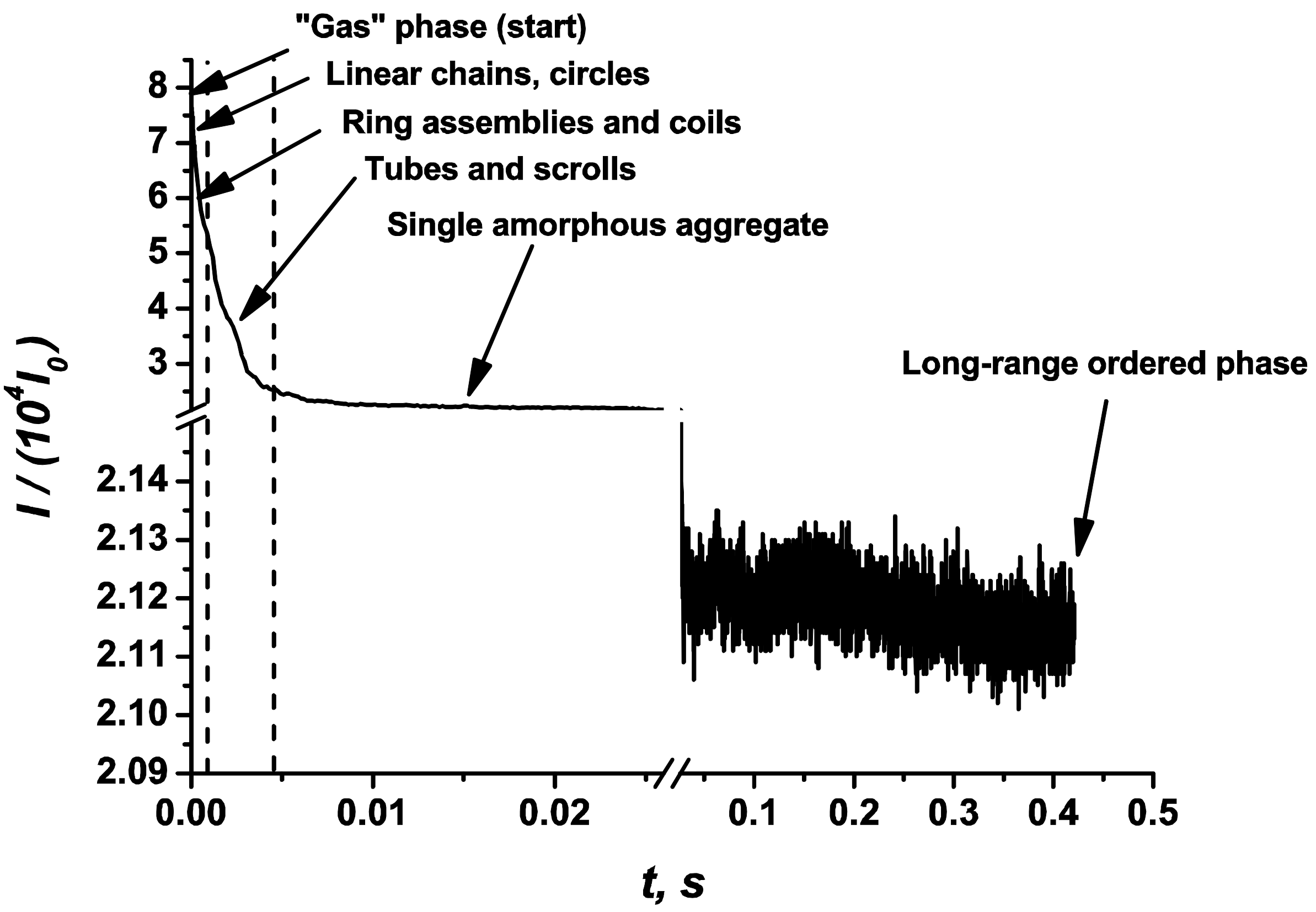

**b)**

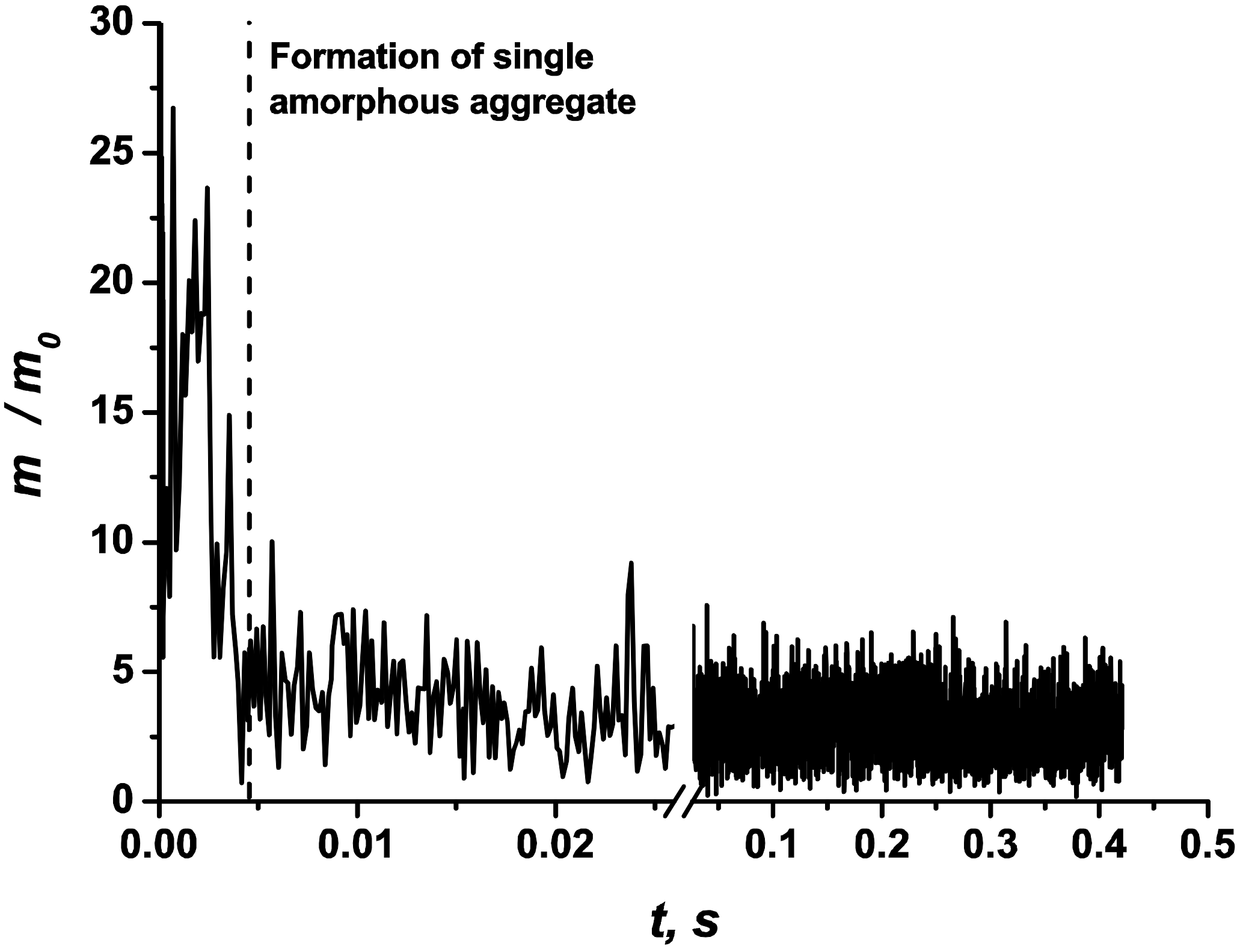

**Fig 1.** The total moment of inertia (a) and the total magnetic moment magnitude (b) time dependencies. The normalized constants are $I_0 = MR^2$ (a) and magnetic moment of single nanoparticles (b). The ranges of the different phases are shown.

a)

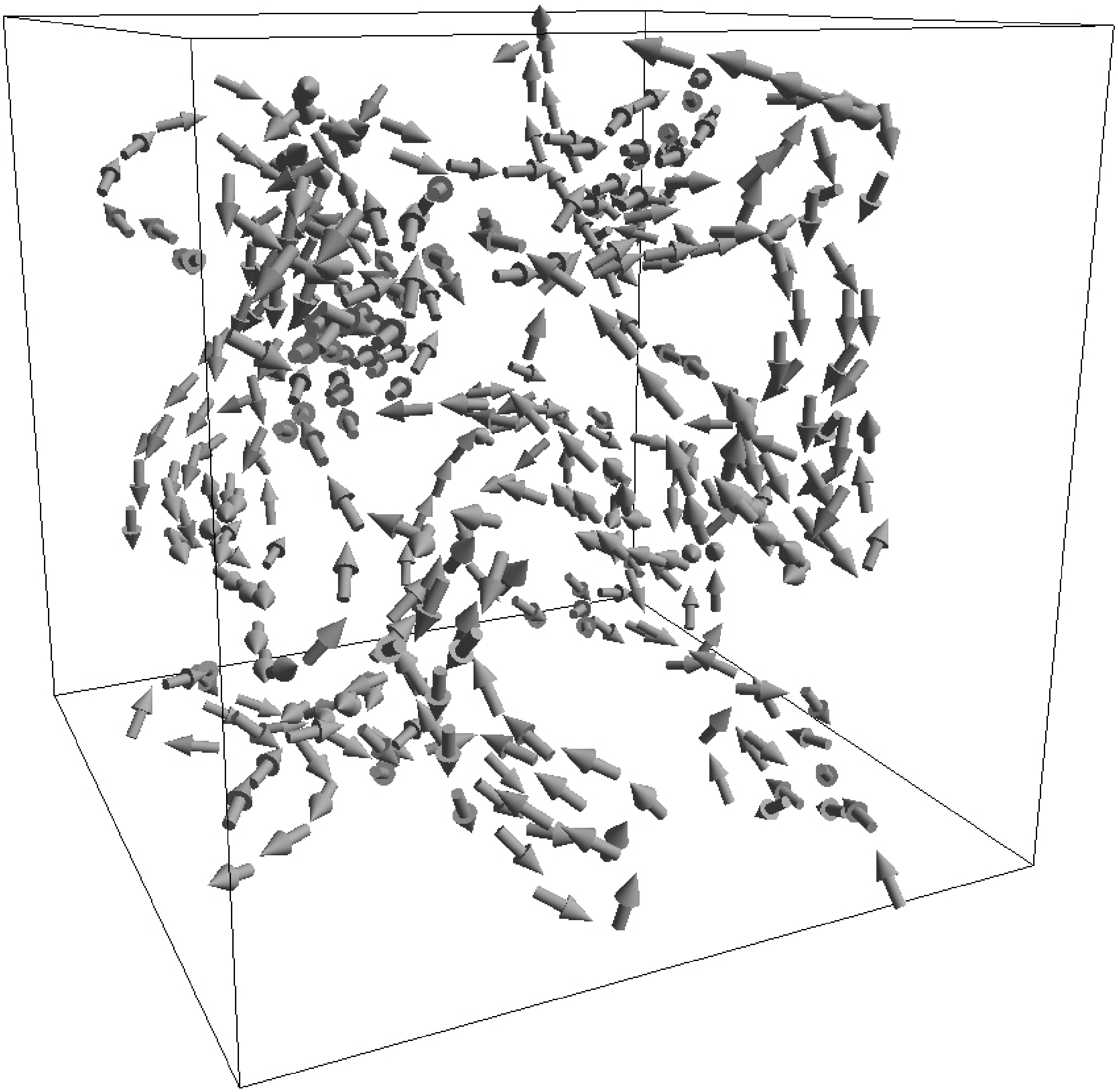

b)

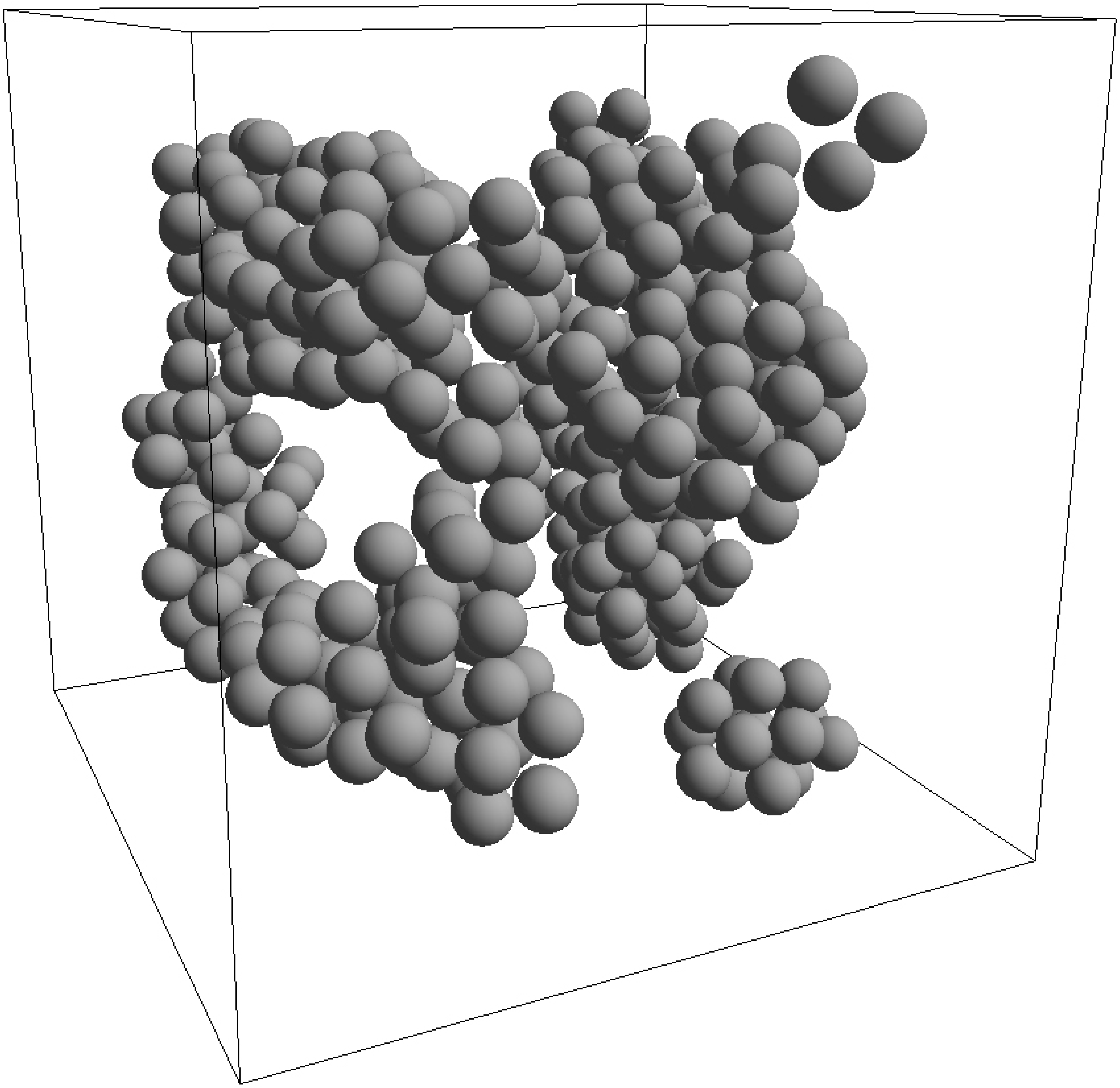

c)

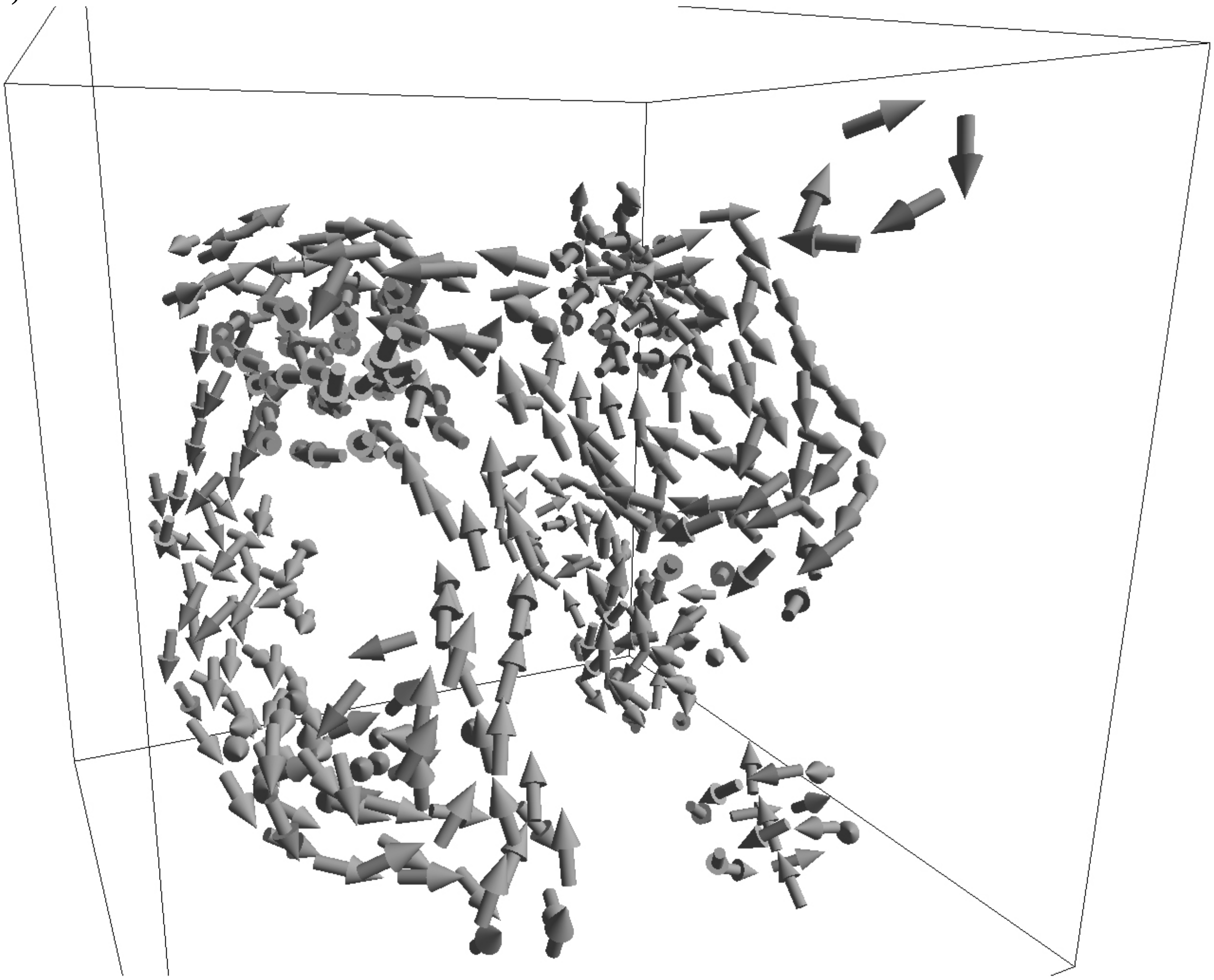

d)

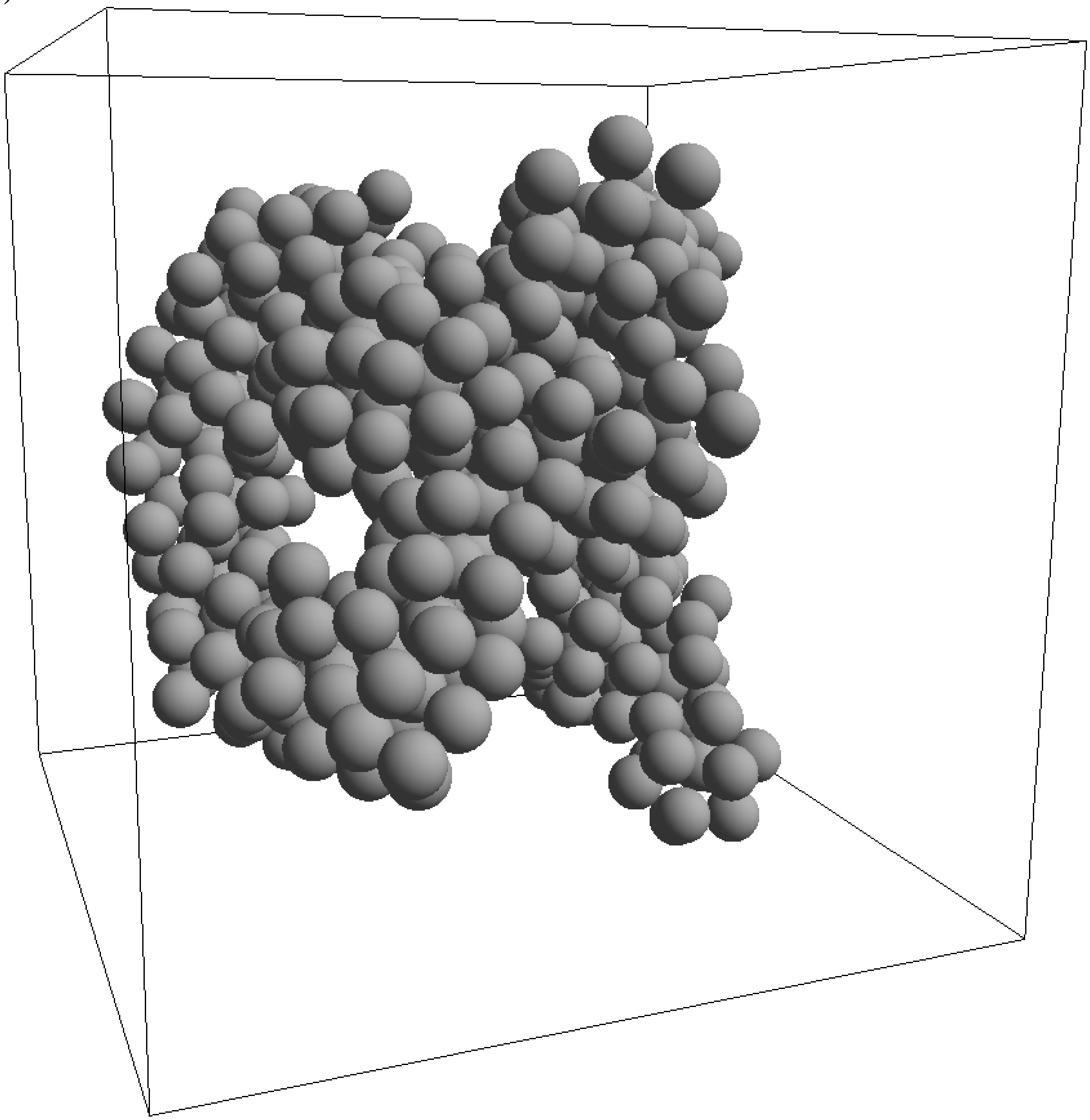

e)

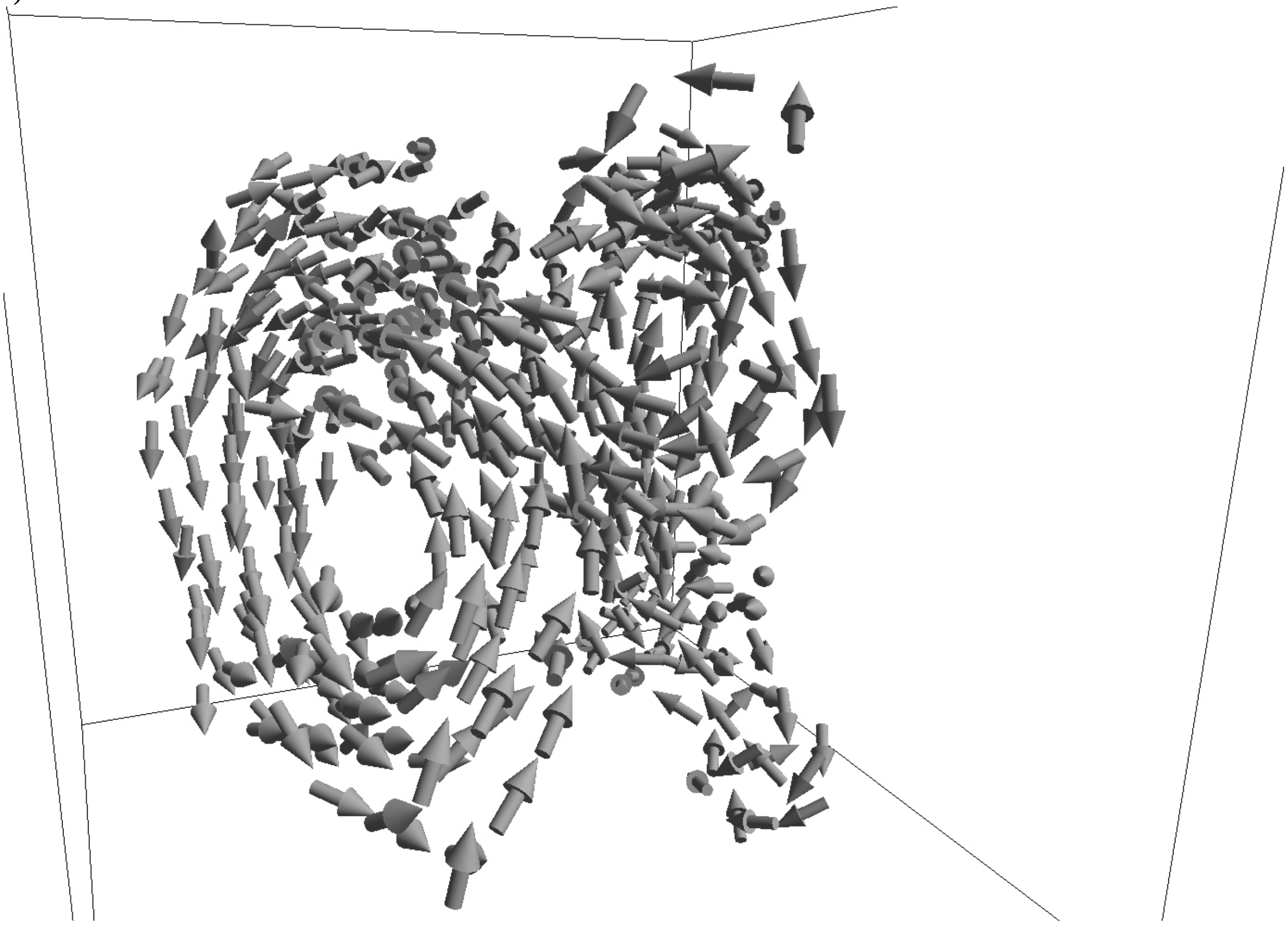

f)

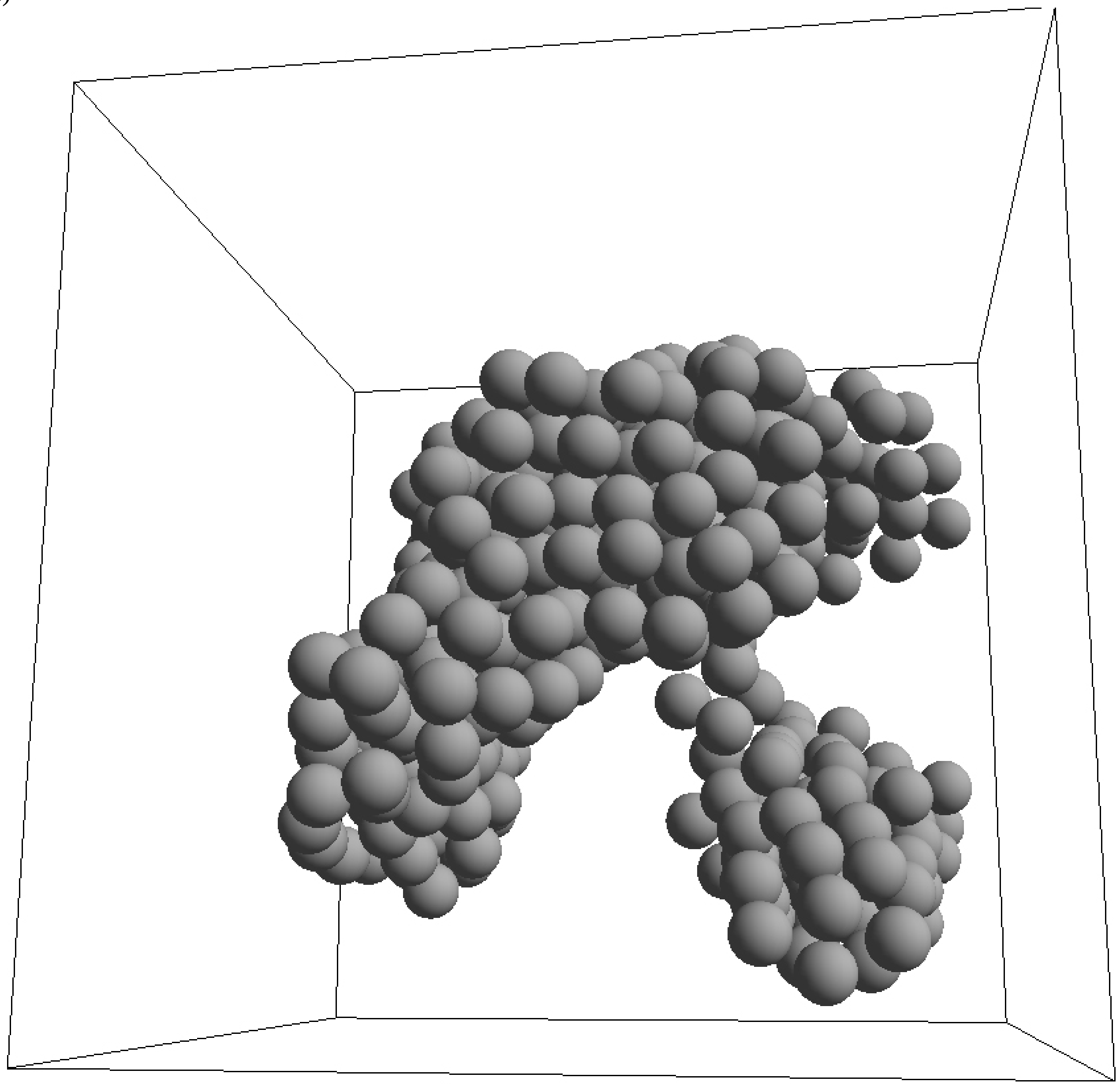

g)

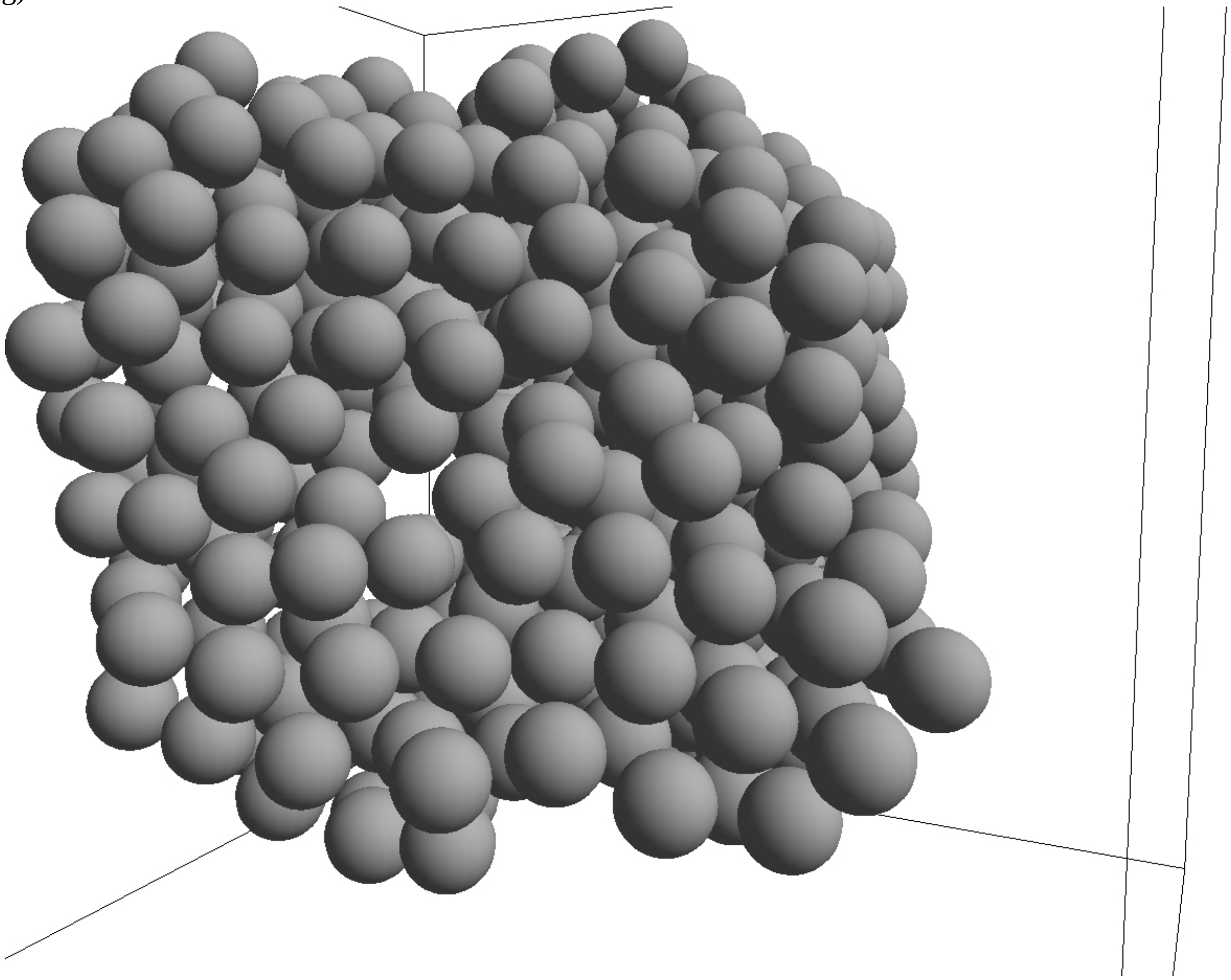

**Fig 2.** The microscopic structure of the ferrofluid at the liquid aggregate formation stage at the time stamps: 0.2 (a), 0.9 (b,c), 1.7 (d,e), 2.4 (f) and 2.2 (g) ms. The particles (b,d,f,g) and magnetic moment (a,c,e) distribution were shown.

a)

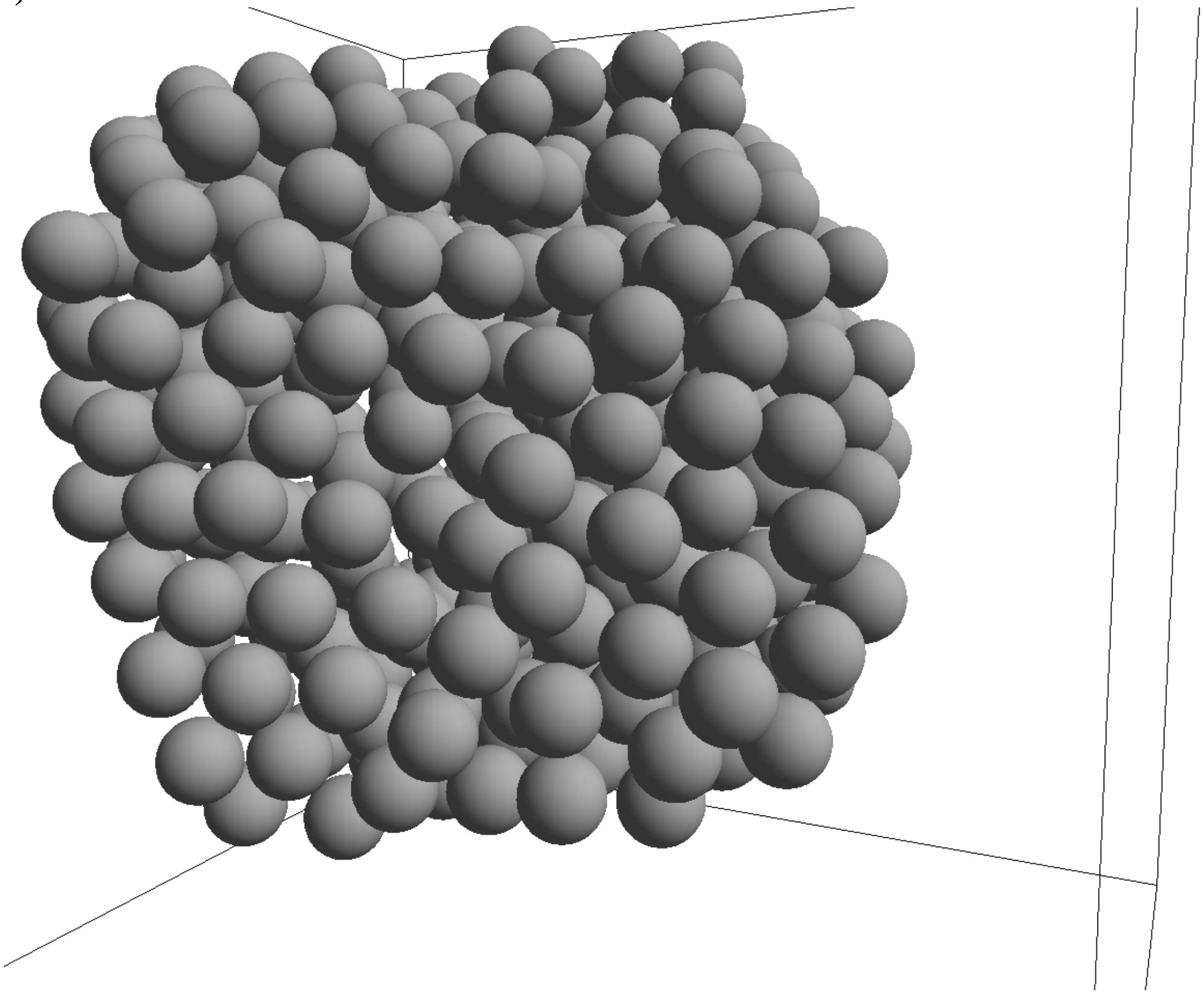

b)

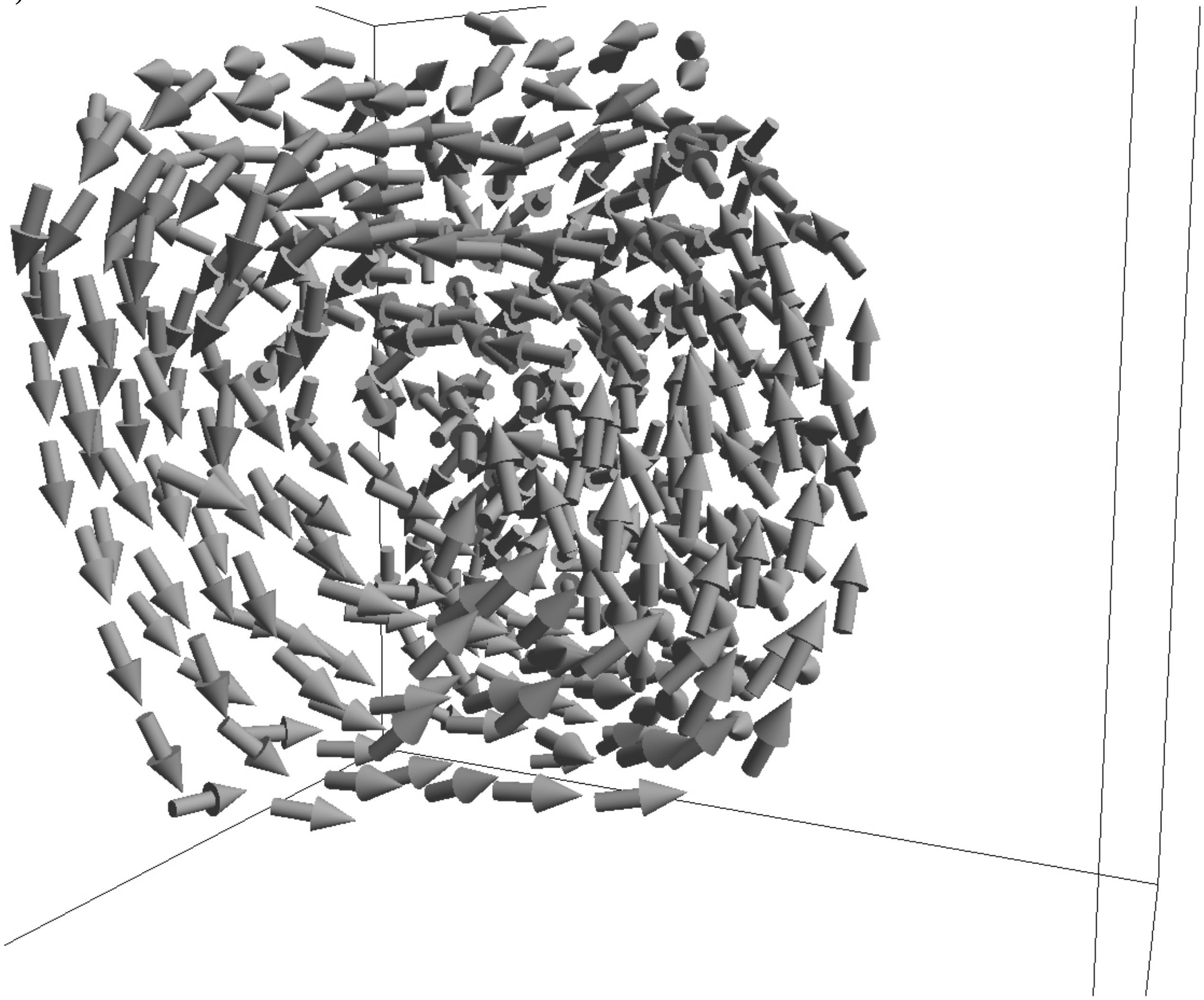

c)

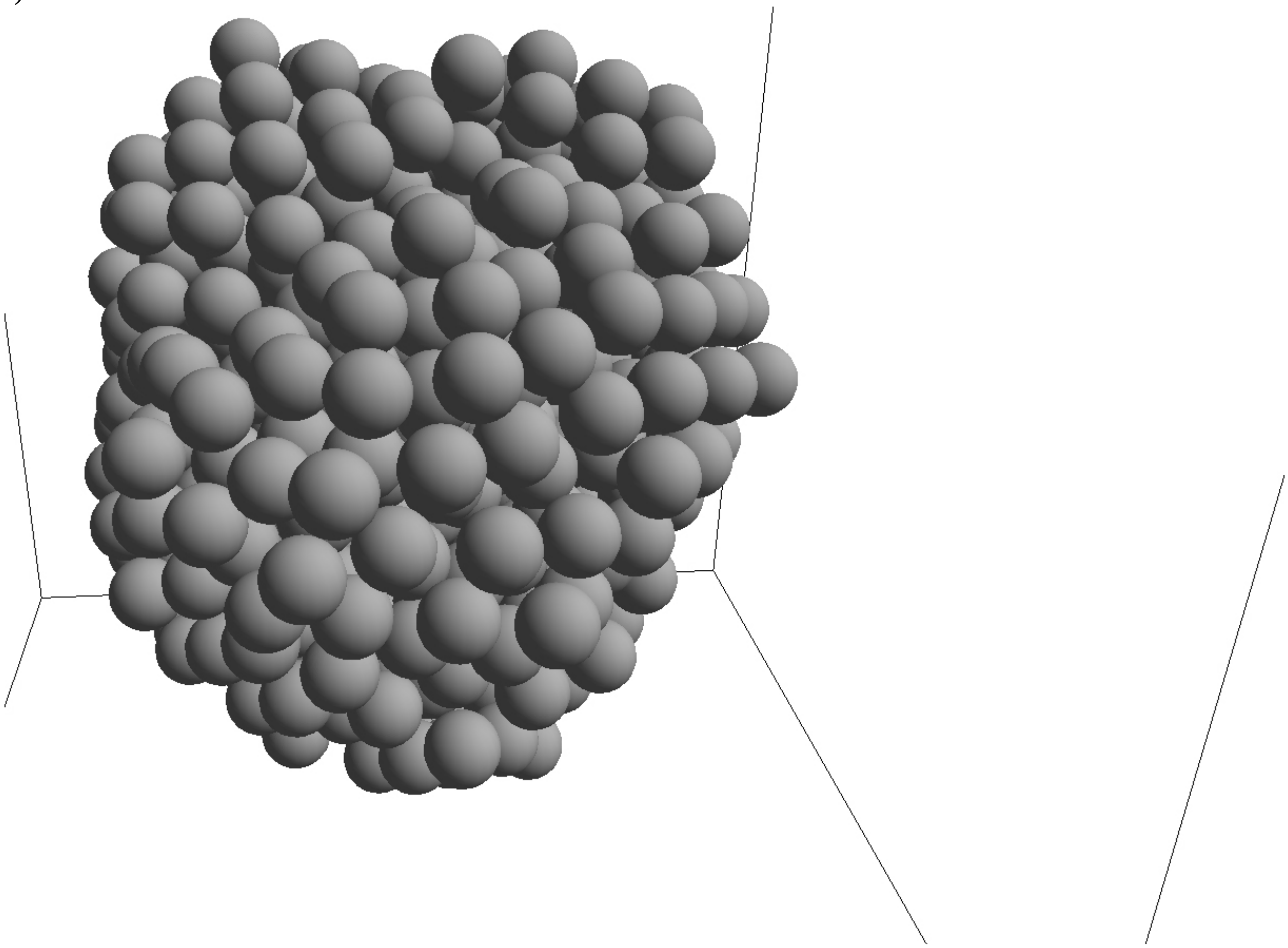

d)

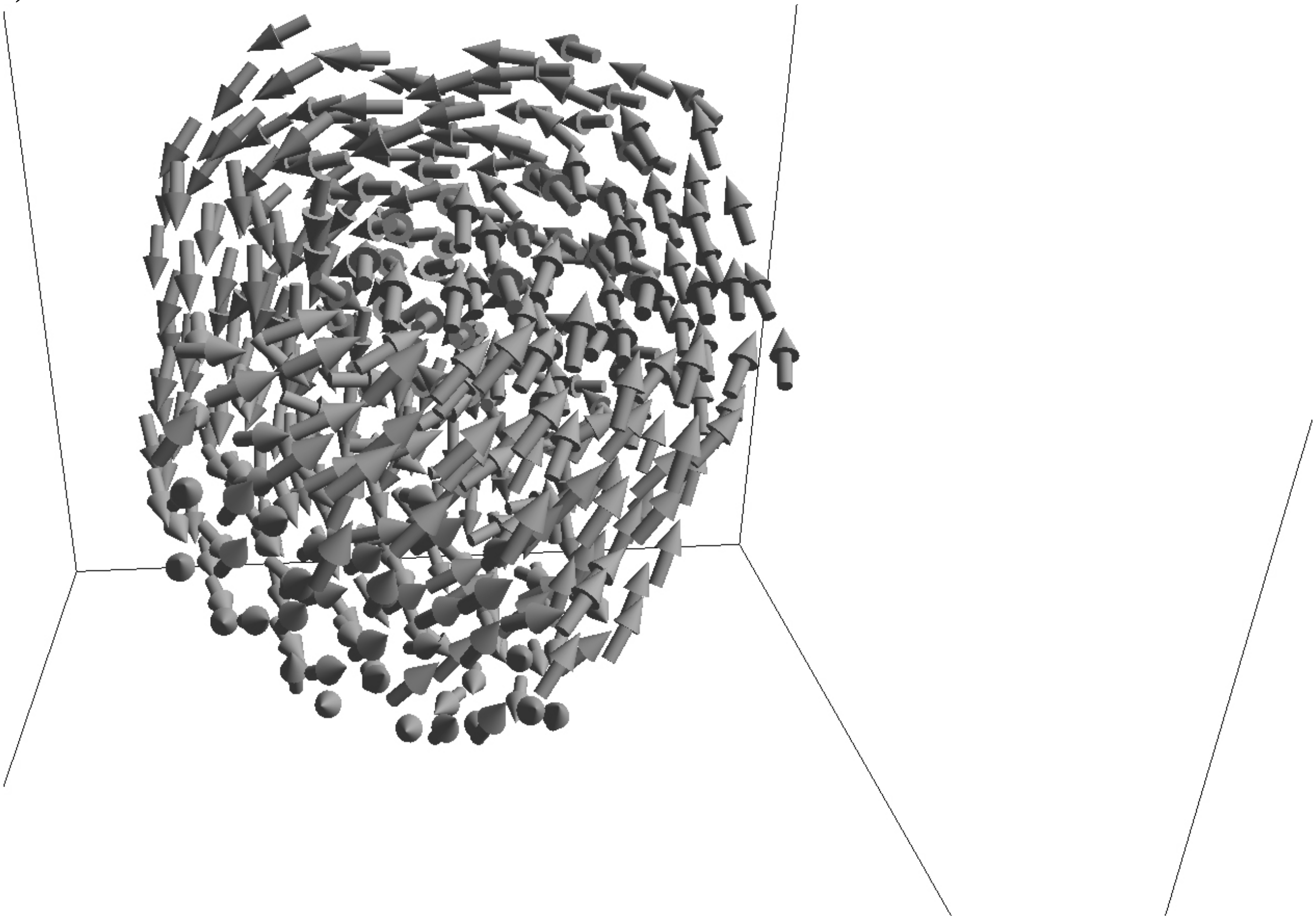

e)

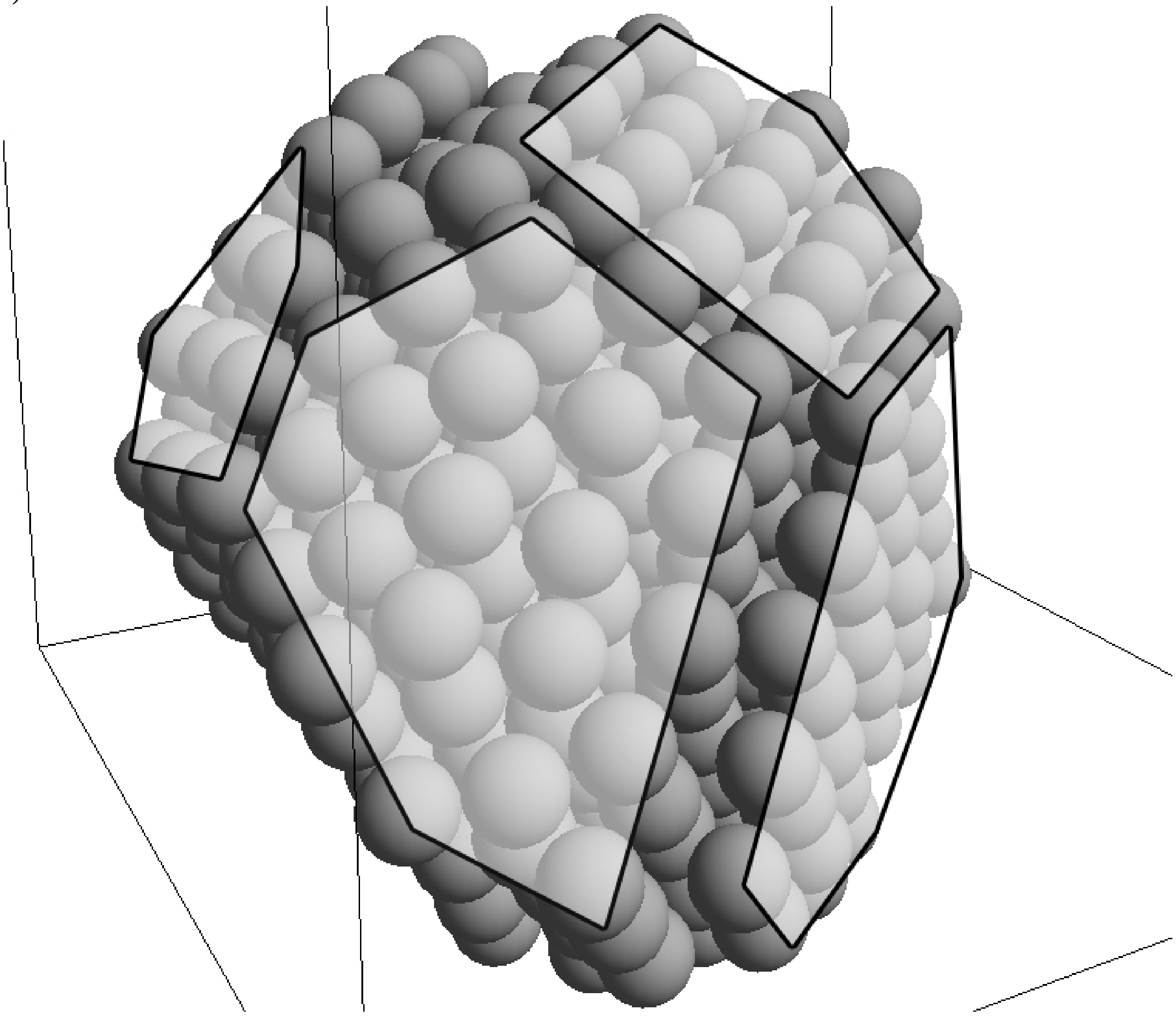

f)

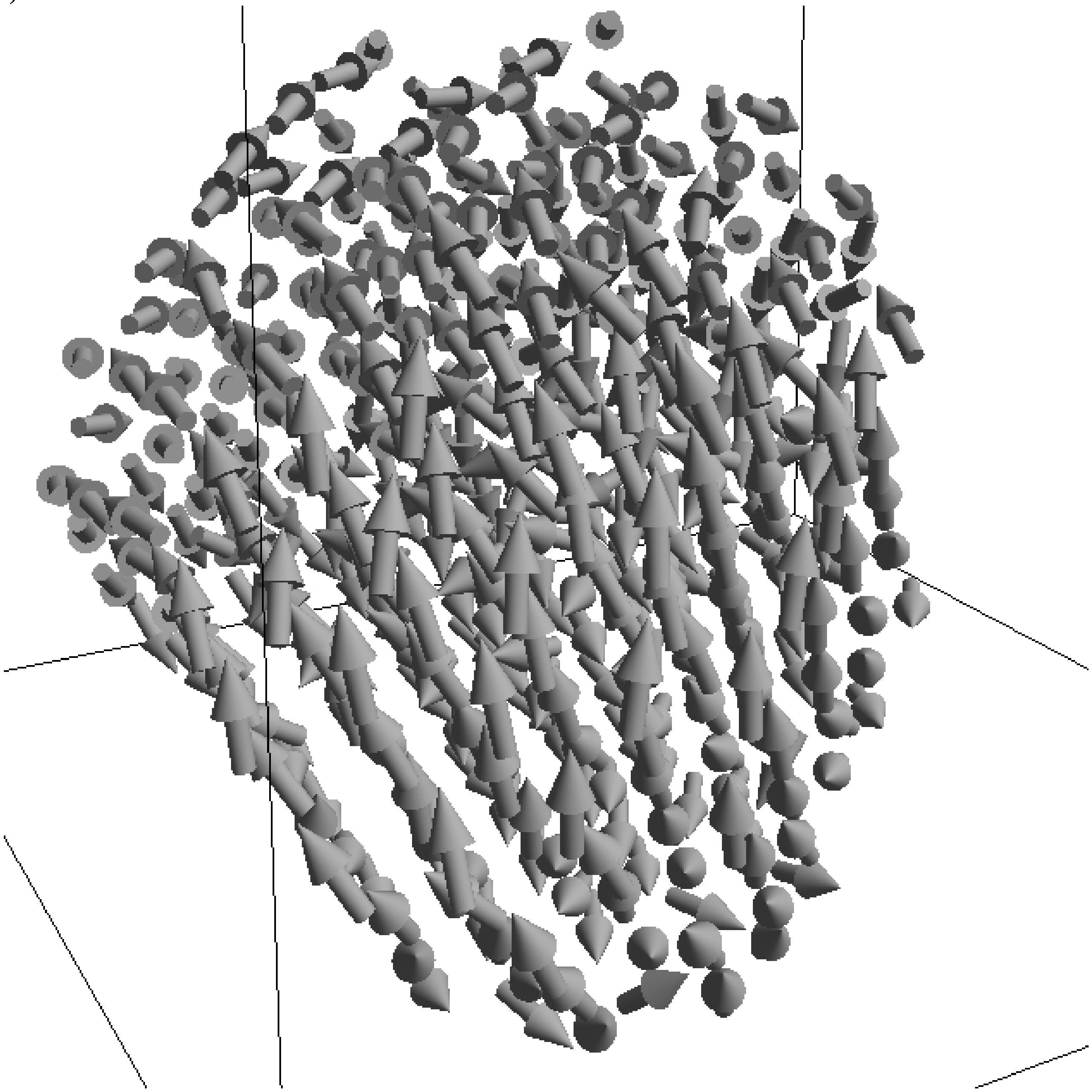

g)

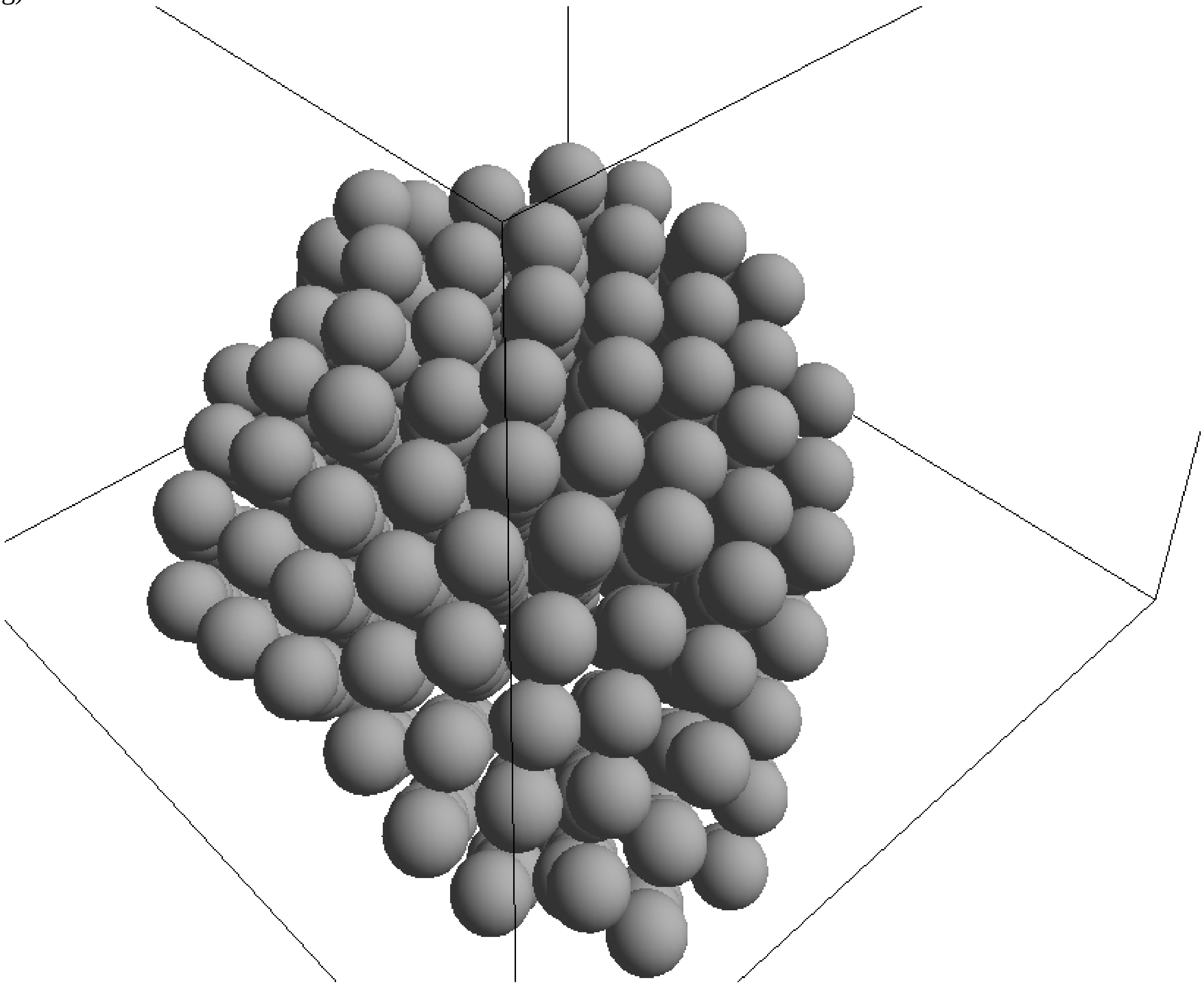

**Fig 3.** The microscopic structure of the ferrofluid at the long-range ordered aggregate formation stage at the time stamps: 4.6 (a,b), 15.7 (c,d) and 421.2 (f,g,e) ms. The particles (a,c,g,e) and magnetic moment (b,d,f) distribution were shown. The crystal faceting is highlighted (e).